\documentclass[twocolumn,amsmath,amssymb,prl,superscriptaddress]{revtex4-1}

\usepackage{graphicx} 
\usepackage{dcolumn} 
\usepackage{bm} 
\makeatletter

\newcommand{\Rmnum}[1]{\expandafter\@slowromancap\romannumeral #1@}
\makeatother

\usepackage{graphicx}
\usepackage{dcolumn}
\usepackage{bm}

\begin{document}

\title{{\it Ab initio} determination of  excitation energies and magnetic couplings in correlated, quasi two-dimensional iridates}

\author{Vamshi M. Katukuri}
\affiliation{Institute for Theoretical Solid State Physics, IFW Dresden,
                   Helmholtzstrasse 20, 01069 Dresden, Germany}
\author{Hermann Stoll}
\affiliation{Institute for Theoretical Chemistry,
                   Universit\"{a}t Stuttgart, Pfaffenwaldring 57, 70550 Stuttgart, Germany}
\author{Jeroen van den Brink}
\affiliation{Institute for Theoretical Solid State Physics, IFW Dresden,
                   Helmholtzstrasse 20, 01069 Dresden, Germany}
\author{Liviu Hozoi}
\affiliation{Institute for Theoretical Solid State Physics, IFW Dresden,
                   Helmholtzstrasse 20, 01069 Dresden, Germany}

\date{\today}


\begin{abstract}
To determine the strength of essential electronic and magnetic interactions in the iridates Sr$_2$IrO$_4$ and Ba$_2$IrO$_4$ --\,potential platforms for high-temperature superconductivity\,-- we use many-body techniques from wavefunction-based electronic-structure theory. Multiplet physics, spin-orbit interactions, and Ir--O hybridization are all treated on equal footing, fully {\it ab initio}. Our calculations put the lowest $d$--$d$ excitations of Sr$_2$IrO$_4$/Ba$_2$IrO$_4$ at 0.69/0.64 eV, substantially lower than in isostructural cuprates. Charge-transfer excitations start at 3.0/1.9 eV and the magnetic nearest-neighbor exchange coupling is 51/58 meV.  Available experimental results are fully consistent with these values, which strongly constrains the parametrization of effective iridate Hamiltonians.
\end{abstract}

\maketitle

{\it Introduction ---}
Electronic correlations in quasi two-dimensional (2D) materials, especially transition metal compounds, have been a central issue in condensed matter physics since the discovery of the cuprate superconductors. While electron-electron interactions are very substantial in $3d$ transition metal compounds~\cite{Imada_3doxides_RevModPhys_98}, they become progressively weaker when going to heavier transition metal elements, i.e., $4d$ and $5d$ systems, as the $d$ orbitals become more and more extended. Interestingly, the relativistic spin-orbit coupling (SOC) follows the opposite trend -- it increases progressively. In $5d$ transition metal compounds, e.g., iridates, the intriguing situation arises where these interactions meet on the same energy scale. The interplay between crystal-field effects, local multiplet physics, SOC's, and intersite hopping that subsequently emerges has opened up a new window of interest in strongly correlated compounds, offering novel types of correlated ground states (GS's) and excitations.

The more exotic examples are possible topological states in iridates, such as a topological Mott insulator~\cite{IrO_TIs_balents_10,IrO_TIs_yang_10}, a Weyl semimetal and axion insulator~\cite{IrO_weyl_savrasov_11}, and the possible realization of the long-sought Kitaev model with bond-dependent spin-spin interactions~\cite{IrO_kitaev_jackeli_09,IrO_kitaev_jackeli_10,IrO_kitaev_trebst_11}. The 2D square-lattice iridates such as Sr$_2$IrO$_4$ and Ba$_2$IrO$_4$ are on the other hand appealing because of their perceived structural and magnetic similarity to La$_2$CuO$_4$, the mother compound of the cuprate high-T$_{\rm c}$ (HTC) superconductors, which has promoted them to novel platforms on which HTC superconductivity may be designed~\cite{IrO_mott_kim_08,IrO_mott_arima_09,IrO_rixs_kim_11,IrO_HTC_senthil_11}. To put such considerations on a firm footing it is essential to quantify the different coupling strengths and energy scales, as they for instance appear in effective Hamiltonian descriptions of these correlated systems~\cite{IrO_HTC_senthil_11,IrO_mott_yunoki_10,IrO_mott_arita_11}.

Whereas the relevant single-particle hopping integrals can be extracted from tight-binding fits of density-functional-based band structures \cite{IrO_mott_kim_08,IrO_mott_yunoki_10,IrO_mott_arita_11,IrO_mott_jin_09}, for a well-founded evaluation of interactions strengths and excitation energies from first principles one has to rely on wavefunction many-body approaches~\cite{CuO_J_Illas_00,CuO_dd_hozoi_11,reimers_hozoi_fulde}. Here, we present the outcome and discuss the implications of {\it ab initio} many-body calculations for the two so called 214 iridates, Sr$_2$IrO$_4$ and Ba$_2$IrO$_4$. We find that for Sr/Ba 214 the lowest $d$--$d$ excitations of the magnetic GS doublet occur at 0.69/0.64 eV, substantially lower than in isostructural cuprates, where these excitations start at 1.0--1.5 eV~\cite{CuO_dd_hozoi_11}. The charge-transfer (CT) continuum of the 2D iridates commences at 3.0/1.9 eV, with the lowest CT excitations confined to the out-of-plane oxygen ligands.  It is quite remarkable that in spite of the very different Ir--O--Ir bond angles in the two materials, the strength of the magnetic nearest-neighbor (NN) exchange coupling is rather similar, 51 and 58 meV, respectively. This perhaps unexpected result is a direct consequence of the strong SOC and it sets the 2D iridates apart from the $3d$ transition metal systems, in which superexchange couplings in general depend extremely sensitively on these geometrical factors. 

\begin{figure}[b!]
\centering
\includegraphics*[angle=0,scale=0.39]{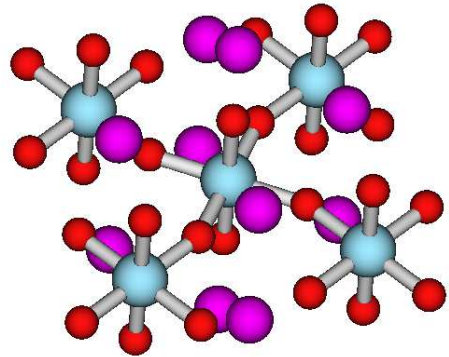}
\caption{ Sketch of the 5-octahedra cluster used for the calculations of crystal-field and CT excitations in Sr$_2$IrO$_4$. Ir, O, and Sr ions are shown in light blue, red, and pink, respectively. In Ba$_2$IrO$_4$, the Ir--O--Ir bonds are straight \cite{IrO_str214_crawford94,IrO_str214_okabe11}.
}
\label{fig:structures}
\end{figure}

{\it Computational method ---}
For our calculations we rely on multiconfiguration self-consistent-field (MCSCF) and multireference configuration-interaction (MRCI) techniques from modern quantum chemistry \cite{book_QC_00}. Multiorbital and multiplet physics, SOC's, and O $2p$ to Ir $5d$ CT effects are all treated on equal footing, fully {\it ab initio}. Since spin-orbit interactions can be switched on and off by hand, we can filter out and explicitly determine the effect of  SOC on the electronic structure and excitations. That the quantum-chemical calculations offer insights into the electronic structure of correlated solids that go substantially beyond standard density-functional approaches 
is well-established by now, e.g., for the 
2D Cu oxides \cite{CuO_J_Illas_00,CuO_dd_hozoi_11,reimers_hozoi_fulde}.
The magnetic exchange constants \cite{CuO_J_Illas_00} and $d$--$d$ excitations \cite{CuO_dd_hozoi_11} obtained in those systems by quantum-chemical calculations are in excellent agreement with the experiment. Further, spin renormalization effects on the cuprate quasiparticle bands have also been determined~\cite{reimers_hozoi_fulde}.
Since the quantum-chemical methodology does not use any {\it ad hoc} parameters, such calculations are very different from, e.g., hybrid schemes \cite{IrO_mott_arita_11} based on density-functional theory (DFT) and dynamical mean-field theory (DMFT)  and provide information on the electronic structure and excitations that is complementary to 
for instance 
a DFT+DMFT approach
 \cite{IrO_mott_arita_11}.

In the spirit of modern multiscale electronic-structure approaches, a given region of the extended crystal is treated by advanced quantum-chemical many-body techniques while the remaining part of the solid is modeled at a more elementary level \cite{CuO_J_Illas_00,CuO_dd_hozoi_11,reimers_hozoi_fulde}. 
For each of the two compounds, we first designed 5-octahedra clusters (see Fig.\,1) to address the nature of the GS electron configuration and of the lowest crystal-field (CF) and CT excited states. The ten Sr or Ba ions next to the central IrO$_6$ octahedron are also included in these clusters. 
The remaining part of the crystal is modeled as a large array of point charges (PC's) which reproduce the Madelung potential in the cluster region.
For the GS calculations, the orbitals within each finite cluster are variationally optimized at the MCSCF level.
All Ir $t_{2g}$ functions are included in the active orbital space \cite{book_QC_00}, i.e., all possible electron occupations are allowed within the $t_{2g}$ set of orbitals. On-site and CT excitations are afterwards computed just for the central IrO$_6$ octahedron while the occupation of the NN Ir valence shells is held frozen as in the GS configuration. We analyze then the correlated electronic structure of these compounds at the MCSCF and MRCI levels of theory, both with and without SOC's \cite{WK88,SOC_molpro}. To extract the NN superexchange $J$, we employed PC-embedded 8-octahedra clusters. For those clusters, high-spin and low-spin magnetic configurations are computed for two active Ir sites. To reduce the complexity of the problem and the computational effort, the six adjacent Ir$^{4+}$ ions are modeled as closed-shell Pt$^{4+}$ $t_{2g}^6$ species.

All calculations were performed with the {\sc molpro} quantum-chemical software \cite{molpro_brief}. We used energy-consistent relativistic pseudopotentials for Ir, Sr, and Ba~\cite{ECP_Stoll_2,ECP_Stoll_1} and Gaussian-type basis functions from the {\sc molpro} library. Basis sets of quadruple-zeta quality \cite{book_QC_00} were applied for the valence shells of the central Ir ion and triple-zeta basis sets for the ligands of the central octahedron and the NN Ir sites. For the central Ir ions we also used two polarization $f$ functions. For farther ligands around the NN Ir sites we applied minimal basis sets. All occupied shells of the Sr$^{2+}$ and Ba$^{2+}$ ions were incorporated in the large-core pseudopotentials \cite{ECP_Stoll_1} and each of the Sr $5s$ and Ba $6s$ orbitals was described by a single contracted Gaussian function. The ligand bridging the two active Ir sites in the 8-octahedra clusters was described with  quintuple-zeta valence basis sets and four polarization $d$ functions. 

\begin{table}[!t]
\caption{Splittings within the Ir $t_{2g}$ shells in Sr$_2$IrO$_4$ and Ba$_2$IrO$_4$,
with and without SOC.
MCSCF and MRCI results for 5-octahedra embedded clusters, hole representation. 
}
\begin{ruledtabular}
\begin{tabular}{llll}
Hole configuration    &\multicolumn{3}{l}{Relative energies (eV)}    \\
\hline
\\
Sr$_2$IrO$_4$ : \\
$t_{2g}^1$ -- no SOC, $S\!=\!1/2$
                      &$d_{xy}^1$       &$d_{xz}^1$       &$d_{yz}^1$ \\
MCSCF                 &0                &0.12             &0.12       \\
MRCI                  &0                &0.06             &0.06       \\
$t_{2g}^1$ -- SOC     &$j=1/2$          &$j=3/2$          &           \\
MRCI+SOC              &0                &0.65\,--\,0.81   &           \\
\\
Ba$_2$IrO$_4$ : \\
$t_{2g}^1$ -- no SOC, $S\!=\!1/2$
                      &$d_{xz}^1$       &$d_{yz}^1$       &$d_{xy}^1$ \\
MCSCF                 &0                &0                &0.19       \\
MRCI                  &0                &0                &0.27       \\
$t_{2g}^1$ -- SOC     &$j=1/2$          &$j=3/2$          &           \\
MRCI+SOC              &0                &0.69\,--\,0.82   &           \\
\end{tabular}
\end{ruledtabular}
\label{dd_t2g}
\end{table}

{\it Splittings of Ir $t_{2g}$ levels ---}
For the (formal) Ir$^{4+}$ $t_{2g}^5$ valence electron configuration, the strong SOC tends to split off a low energy $j=1/2$ doublet. Much attention has been recently paid to obtaining the size of this splitting, the most basic information on the SOC strength $\lambda$~\cite{IrO_mott_kim_08,IrO_mott_jin_09,IrO_rixs_ament_11}, and the nature and size of the intersite spin-exchange interactions  in a number of different iridates \cite{IrO_kitaev_jackeli_10,IrO_kitaev_jackeli_09,IrO_kitaev_trebst_11,IrO_kitaev_heisenberg_kimchi_11,IrO_rixs_kim_11,NaIrO_mgn_kim_11,NaIrO_mgn_yu_11}. As concerns the magnetic interactions, it has been pointed out that fine details such as the size of the splitting(s) within the valence Ir $t_{2g}$ shell can be extremely relevant.  In the layered honeycomb systems, for instance, the trigonal CF splittings appear to strongly influence the nature of the magnetic GS \cite{NaIrO_mgn_kim_11,NaIrO_mgn_yu_11}.

Whereas the crystal structures of the 214 iridates are similar to that of La$_2$CuO$_4$, an important difference between the Sr and Ba 214 systems is that in the former material the IrO$_6$ octahedra are rotated about the $c$ axis such that the  in-plane Ir--O--Ir angles are not 180$^{\circ}$ while in the latter the  Ir--O--Ir bonds are straight.
In both Sr$_2$IrO$_4$ and Ba$_2$IrO$_4$, the apical Ir--O bonds are longer than
the in-plane bonds, 2.06 vs.~1.98 \AA \ in  Sr$_2$IrO$_4$ and 2.15 vs.~2.01 \AA \ 
in Ba$_2$IrO$_4$ \cite{IrO_str214_crawford94,IrO_str214_okabe11}.
Results for the splittings within the Ir $t_{2g}$ shells in the two materials are
listed in Table\,I.
Interestingly, the calculations in which the SOC is switched off show that, in the electron representation,
the lowest, doubly occupied $t_{2g}$ levels are $d_{xz}$ and $d_{yz}$ in Sr$_2$IrO$_4$
and $d_{xy}$ in Ba$_2$IrO$_4$.
By MRCI calculations, the splittings within the $t_{2g}$ shell are 0.06 and 0.27 eV,
respectively (see Table\,I).
We included in the MRCI treatment single and double excitations from the Ir $t_{2g}$ 
and all O $2p$ orbitals at the central octahedron.

Test calculations in which for the crystal structure of Sr$_2$IrO$_4$ we replaced the
Sr ions by Ba show that the $d_{xz}$ and $d_{yz}$ levels are still the lowest, as for 
the set of results listed in the upper part of Table\,I.
Also, if we use the crystal structure of Ba$_2$IrO$_4$ but replace Ba by Sr we still
find that the lowest-energy $t_{2g}$ level is $d_{xy}$, as for the results given
in the lower part of the table.
This shows that our findings of a different energy order of the Ir $t_{2g}$ levels
for Sr and Ba 214 in absence of SOC are strictly related to local Ir--O bonding.
Nevertheless, when SOC is accounted for, the splittings between the local doublet
$j\!=\!1/2$ and quartet $j\!=\!3/2$ states are about the same in the two compounds.
As mentioned above, these calculations were performed on embedded 5-octahedra clusters.
In a nonrelativistic picture, the five Ir $S\!=\!1/2$ sites can couple to sextet, quartet,
and doublet states.
All those different spin configurations entered the spin-orbit calculations so that the energy windows 
of 0.65\,--\,0.81 eV in Sr$_2$IrO$_4$ and 0.69\,--\,0.82 eV
in Ba$_2$IrO$_4$ actually include a total number of 64 spin-orbit final states~\cite{footnote_SO_96}.
A simplified picture can be obtained by replacing the Ir$^{4+}$ $t_{2g}^5$ NN's by 
closed-shell Pt$^{4+}$ $t_{2g}^6$ ions. 
The energies of the two components of the $j\!=\!3/2$ quartet, split off by the tetragonal
field, are in that case 0.69 and 0.73 in Sr$_2$IrO$_4$ and 0.64 and 0.71 in Ba$_2$IrO$_4$.
These results for the energy separation between the $j\!=\!1/2$ and $j\!=\!3/2$ states imply an 
effective spin-orbit coupling parameter $\lambda$ for Sr/Ba 214 of 0.46/0.43 eV, which is surprisingly close to values of 0.39--0.50 eV extracted for Ir$^{4+}$ impurities from electron spin resonance measurements~\cite{Andlauer76,Su09}.

\begin{table}[!t]
\caption{
Nearest-neighbor exchange coupling $J$ in Sr$_2$IrO$_4$ and Ba$_2$IrO$_4$.
Negative/positive values denote FM/AF interactions.
MCSCF and MRCI results for clusters with two active magnetic centers, see text.
}
\begin{ruledtabular}
\begin{tabular}{lrr}
$J$ (meV)             &Sr$_2$IrO$_4$    &Ba$_2$IrO$_4$    \\
\hline
\\
MCSCF                 &$-19.2$          &15.4  \\
MCSCF+SOC             &$28.8$           &35.7  \\
MRCI+SOC              &$51.3$           &58.0  \\
\end{tabular}
\end{ruledtabular}
\label{dd_t2g}
\end{table}

{\it CF and CT excitations ---}
MRCI calculations without SOC put the CF excitations from the $t_{2g}$ to the $e_g$ levels, $t_{2g}^5\!\rightarrow\!t_{2g}^4e_g^1$ and $t_{2g}^5\!\rightarrow\!t_{2g}^3e_g^2$, between 2.8\,--\,6 eV in Sr$_2$IrO$_4$ and between 2.0\,--\,5.0 eV in Ba$_2$IrO$_4$.
The lowest O $2p$ to Ir $5d$ CT states in MRCI calculations without SOC are at 3.6 eV 
above the GS in Sr$_2$IrO$_4$ and at 1.7 eV with respect to the GS in 
Ba$_2$IrO$_4$.
When SOC is switched on, the CF and CT configurations mix and the lowest of
those spin-orbit states require excitation energies of 3.0 eV in Sr$_2$IrO$_4$ and 1.9 eV in
Ba$_2$IrO$_4$.
For Sr$_2$IrO$_4$, those values fit well the resonant inelastic x-ray scattering (RIXS) \cite{IrO_rixs_ishii_11} and optical
absorption data \cite{IrO_optics_moon_06}, with both types of measurements displaying marked spectral features between 2.5 and 8 eV.

For both Sr and Ba 214, the computed CT gaps are relatively small. Nevertheless, the interaction and mixing of non-CT $t_{2g}^5$ and CT $t_{2g}^6\underline{L}$ configurations via SOC is not substantial. It should be noted, however, that a major difference with respect to the 2D $S\!=\!1/2$ cuprates is that in the iridates the lowest CT states have apical hole character, whereas in the 2D Cu oxides the lowest electron-removal states have predominant in-plane O $2p$ character, see, e.g., Ref.~\cite{reimers_hozoi_fulde}.

{\it Magnetic exchange couplings ---}
Our MRCI calculations for the superexchange coupling $J$ include single and double excitations from the Ir $t_{2g}$ and bridging-ligand $2p$ orbitals. Quantum chemical calculations of magnetic couplings in terms of restricted configuration-interaction expansions involving the $d$ orbitals at the active magnetic sites and $p$ functions of the bridging ligand were previously reported, e.g., for La$_2$CuO$_4$ \cite{CuO_J_NOCI_03}, providing $J$'s that are in good 
agreement with experiment. All possible occupations were allowed within the set of $t_{2g}$ orbitals at the two Ir sites in the present MCSCF calculations, which gives rise to nine singlet and nine triplet states in absence of SOC. These MCSCF wavefunctions were expressed in terms of orbitals optimized for an average of singlet and triplet states. All those eighteen states entered the spin-orbit calculations, both at the MCSCF and MRCI levels. 
$J$ then corresponds to 
the difference between the energies of the lowest spin-orbit singlet and triplet state. 
The three components of the lowest $j_{\rm total}\!=\!1$ triplet display in both compounds rather small splittings of not more than 3 meV, which shows that the mapping of our {\it ab initio} results on a symmetric Heisenberg model is controlled. The small effects related to the presence of anisotropic non-Heisenberg terms in the interactions will be the subject of future work. 

The resulting $J$'s, as shown in Table~II, have a clear physical interpretation. Since the splittings within the $t_{2g}$ shell are relatively small, the three different $t_{2g}^5$ configurations (see Table~I) appear with similar weights in the spin-orbit GS wavefunctions. Careful analysis of the wavefunctions for clusters with two active Ir sites however shows that the different order of the $t_{2g}$ levels in Sr and Ba 214 gives rise to subtle differences with respect to the superexchange interactions. For corner-sharing plaquettes or octahedra and $t_{2g}$ active orbitals, the Anderson  superexchange implies $d$--$p$--$d$ $\pi$-type links. For Sr$_2$IrO$_4$, when the SOC is switched off, the relevant hole orbital is the in-plane
$d_{xy}$ component, as shown in Table~I. Due to rotation of the IrO$_6$ octahedra and bending of the $d_{xy}$--$p_{x/y}$--$d_{xy}$ bond in Sr 214~\cite{IrO_str214_crawford94} the antiferromagnetic (AF) superexchange is not effective and the spin interaction $J$ is actually ferromagnetic (FM), see Table~II.
With SOC, the $t_{2g}$ hole acquires $d_{xz}$ and $d_{yz}$ character too. Because the $d_{xz/yz}$--$p_{z}$--$d_{xz/yz}$ orbital overlap is rather unaffected by bending of the  Ir--O--Ir bond, the $d$--$p$--$d$ superexchange interactions are much more effective in the presence of SOC. Consequently the magnetic exchange $J$ turns AF, both in MCSCF and MRCI (2nd column in Table~II).

Ba$_2$IrO$_4$, on the other hand, lacks a bending of the Ir--O--Ir bond~\cite{IrO_str214_okabe11}. The nature of the hole orbitals in the GS wavefunction in absence of SOC therefore matters less in this case: $J$ is large and AF, also in absence of SOC's (3rd column in Table\,II).
At the MRCI level and including SOC's,
$J$
turns out to
be somewhat larger in Ba$_2$IrO$_4$ as compared to Sr$_2$IrO$_4$, 58.0 vs.~51.3 meV.

{\it Conclusions ---}
Our detailed investigation of the electronic structure of  the 2D $j\!=\!1/2$ Ir oxides Sr$_2$IrO$_4$ and Ba$_2$IrO$_4$,
with {\it ab initio} many-body quantum-chemical calculations,
shows unequivocally that the peak feature observed at about 0.6 eV in the RIXS~\cite{IrO_rixs_kim_11,IrO_rixs_ishii_11} on Sr 214 is directly related  to transitions between Ir $t_{2g}^5$ $j\!=\!1/2$ and $j\!=\!3/2$ states that are in an energy window of  0.65--0.81 eV in Sr 214 and 0.69--0.82 eV in Ba 214.
It is interesting to note that the peak at this energy in optical absorption measurements was instead assigned to an excitation between Hubbard bands~\cite{IrO_optics_moon_06}. 
The experimentally observed ``gap" between the 0.6 eV feature and higher excited states that appear above 2 eV
in Sr 214~\cite{IrO_rixs_ishii_11,IrO_mott_kim_08,IrO_optics_moon_06} is fully consistent with our calculations.  
We find that the Sr and Ba 214 compounds differ significantly with respect to their tetragonal crystal-field splittings. While in Sr 214 the lowest $t_{2g}$ states have $xz$ and $yz$ symmetry, in Ba 214 the lowest $t_{2g}$ levels are the $xy$ orbitals. The corresponding tetragonal splittings are 0.06 eV in Sr$_2$IrO$_4$ and 0.27 eV in Ba$_2$IrO$_4$. The calculations further show that without SOC the distorted Ir--O--Ir bonds in Sr 214 would actually give rise to a  FM nearest-neighbor magnetic exchange $J$. However, the strong SOC reverses the sign of $J$ and causes antiferromagnetism. 
We find that the dominant magnetic exchange interaction is of Heisenberg type and  for Sr 214 has a substantial coupling constant $J\!=\!51$ meV, which compares well with the value of 45 meV estimated from experimental data~\cite{IrO_kitaev_jackeli_09} and with RIXS measurements that indicate $J\!=\!60$ meV for this system~\cite{IrO_rixs_kim_11}. For Ba 214, we find a $J$ that is even somewhat larger, which renders it roughly a factor of 2 smaller than the $J$'s in typical 2D cuprates \cite{CuO_J_Illas_00}. This might in itself be encouraging for a scenario of magnetic mediated superconductivity in doped 214 iridates. However, the fact the lowest charge-transfer states in 214's have instead of planar, apical oxygen character, points at a destabilization of Zhang-Rice singlets and consequently an effective low energy Hamiltonian that is different from the standard $t$-$J$ model of doped cuprates.

{\it Aknowledgements ---} We thank G. Khaliullin for stimulating discussions.
L.~H. acknowledges financial support from the German Research Foundation
(Deutsche Forschungsgemeinschaft, DFG).


\begin{thebibliography}{35}%
\makeatletter
\providecommand \@ifxundefined [1]{%
 \@ifx{#1\undefined}
}%
\providecommand \@ifnum [1]{%
 \ifnum #1\expandafter \@firstoftwo
 \else \expandafter \@secondoftwo
 \fi
}%
\providecommand \@ifx [1]{%
 \ifx #1\expandafter \@firstoftwo
 \else \expandafter \@secondoftwo
 \fi
}%
\providecommand \natexlab [1]{#1}%
\providecommand \enquote  [1]{``#1''}%
\providecommand \bibnamefont  [1]{#1}%
\providecommand \bibfnamefont [1]{#1}%
\providecommand \citenamefont [1]{#1}%
\providecommand \href@noop [0]{\@secondoftwo}%
\providecommand \href [0]{\begingroup \@sanitize@url \@href}%
\providecommand \@href[1]{\@@startlink{#1}\@@href}%
\providecommand \@@href[1]{\endgroup#1\@@endlink}%
\providecommand \@sanitize@url [0]{\catcode `\\12\catcode `\$12\catcode
  `\&12\catcode `\#12\catcode `\^12\catcode `\_12\catcode `\%12\relax}%
\providecommand \@@startlink[1]{}%
\providecommand \@@endlink[0]{}%
\providecommand \url  [0]{\begingroup\@sanitize@url \@url }%
\providecommand \@url [1]{\endgroup\@href {#1}{\urlprefix }}%
\providecommand \urlprefix  [0]{URL }%
\providecommand \Eprint [0]{\href }%
\providecommand \doibase [0]{http://dx.doi.org/}%
\providecommand \selectlanguage [0]{\@gobble}%
\providecommand \bibinfo  [0]{\@secondoftwo}%
\providecommand \bibfield  [0]{\@secondoftwo}%
\providecommand \translation [1]{[#1]}%
\providecommand \BibitemOpen [0]{}%
\providecommand \bibitemStop [0]{}%
\providecommand \bibitemNoStop [0]{.\EOS\space}%
\providecommand \EOS [0]{\spacefactor3000\relax}%
\providecommand \BibitemShut  [1]{\csname bibitem#1\endcsname}%
\let\auto@bib@innerbib\@empty
\bibitem [{\citenamefont {Imada}\ \emph {et~al.}(1998)\citenamefont {Imada},
  \citenamefont {Fujimori},\ and\ \citenamefont
  {Tokura}}]{Imada_3doxides_RevModPhys_98}%
  \BibitemOpen
  \bibfield  {author} {\bibinfo {author} {\bibfnamefont {M.}~\bibnamefont
  {Imada}}, \bibinfo {author} {\bibfnamefont {A.}~\bibnamefont {Fujimori}}, \
  and\ \bibinfo {author} {\bibfnamefont {Y.}~\bibnamefont {Tokura}},\ }\href
  {\doibase 10.1103/RevModPhys.70.1039} {\bibfield  {journal} {\bibinfo
  {journal} {Rev. Mod. Phys.}\ }\textbf {\bibinfo {volume} {70}},\ \bibinfo
  {pages} {1039} (\bibinfo {year} {1998})}\BibitemShut {NoStop}%
\bibitem [{\citenamefont {Pesin}\ and\ \citenamefont
  {Balents}(2010)}]{IrO_TIs_balents_10}%
  \BibitemOpen
  \bibfield  {author} {\bibinfo {author} {\bibfnamefont {D.}~\bibnamefont
  {Pesin}}\ and\ \bibinfo {author} {\bibfnamefont {L.}~\bibnamefont
  {Balents}},\ }\href@noop {} {\bibfield  {journal} {\bibinfo  {journal}
  {Nature Physics}\ }\textbf {\bibinfo {volume} {6}},\ \bibinfo {pages} {376}
  (\bibinfo {year} {2010})}\BibitemShut {NoStop}%
\bibitem [{\citenamefont {Yang}\ and\ \citenamefont
  {Kim}(2010)}]{IrO_TIs_yang_10}%
  \BibitemOpen
  \bibfield  {author} {\bibinfo {author} {\bibfnamefont {B.-J.}\ \bibnamefont
  {Yang}}\ and\ \bibinfo {author} {\bibfnamefont {Y.~B.}\ \bibnamefont {Kim}},\
  }\href {\doibase 10.1103/PhysRevB.82.085111} {\bibfield  {journal} {\bibinfo
  {journal} {Phys. Rev. B}\ }\textbf {\bibinfo {volume} {82}},\ \bibinfo
  {pages} {085111} (\bibinfo {year} {2010})}\BibitemShut {NoStop}%
\bibitem [{\citenamefont {Wan}\ \emph {et~al.}(2011)\citenamefont {Wan},
  \citenamefont {Turner}, \citenamefont {Vishwanath},\ and\ \citenamefont
  {Savrasov}}]{IrO_weyl_savrasov_11}%
  \BibitemOpen
  \bibfield  {author} {\bibinfo {author} {\bibfnamefont {X.}~\bibnamefont
  {Wan}}, \bibinfo {author} {\bibfnamefont {A.}~\bibnamefont {Turner}},
  \bibinfo {author} {\bibfnamefont {A.}~\bibnamefont {Vishwanath}}, \ and\
  \bibinfo {author} {\bibfnamefont {S.}~\bibnamefont {Savrasov}},\ }\href@noop
  {} {\bibfield  {journal} {\bibinfo  {journal} {Phys. Rev. B}\ }\textbf
  {\bibinfo {volume} {83}},\ \bibinfo {pages} {205101} (\bibinfo {year}
  {2011})}\BibitemShut {NoStop}%
\bibitem [{\citenamefont {Jackeli}\ and\ \citenamefont
  {Khaliullin}(2009)}]{IrO_kitaev_jackeli_09}%
  \BibitemOpen
  \bibfield  {author} {\bibinfo {author} {\bibfnamefont {G.}~\bibnamefont
  {Jackeli}}\ and\ \bibinfo {author} {\bibfnamefont {G.}~\bibnamefont
  {Khaliullin}},\ }\href {\doibase 10.1103/PhysRevLett.102.017205} {\bibfield
  {journal} {\bibinfo  {journal} {Phys. Rev. Lett.}\ }\textbf {\bibinfo
  {volume} {102}},\ \bibinfo {pages} {017205} (\bibinfo {year}
  {2009})}\BibitemShut {NoStop}%
\bibitem [{\citenamefont {Chaloupka}\ \emph {et~al.}(2010)\citenamefont
  {Chaloupka}, \citenamefont {Jackeli},\ and\ \citenamefont
  {Khaliullin}}]{IrO_kitaev_jackeli_10}%
  \BibitemOpen
  \bibfield  {author} {\bibinfo {author} {\bibfnamefont {J.}~\bibnamefont
  {Chaloupka}}, \bibinfo {author} {\bibfnamefont {G.}~\bibnamefont {Jackeli}},
  \ and\ \bibinfo {author} {\bibfnamefont {G.}~\bibnamefont {Khaliullin}},\
  }\href {\doibase 10.1103/PhysRevLett.105.027204} {\bibfield  {journal}
  {\bibinfo  {journal} {Phys. Rev. Lett.}\ }\textbf {\bibinfo {volume} {105}},\
  \bibinfo {pages} {027204} (\bibinfo {year} {2010})}\BibitemShut {NoStop}%
\bibitem [{\citenamefont {Reuther}\ \emph {et~al.}(2011)\citenamefont
  {Reuther}, \citenamefont {Thomale},\ and\ \citenamefont
  {Trebst}}]{IrO_kitaev_trebst_11}%
  \BibitemOpen
  \bibfield  {author} {\bibinfo {author} {\bibfnamefont {J.}~\bibnamefont
  {Reuther}}, \bibinfo {author} {\bibfnamefont {R.}~\bibnamefont {Thomale}}, \
  and\ \bibinfo {author} {\bibfnamefont {S.}~\bibnamefont {Trebst}},\ }\href
  {\doibase 10.1103/PhysRevB.84.100406} {\bibfield  {journal} {\bibinfo
  {journal} {Phys. Rev. B}\ }\textbf {\bibinfo {volume} {84}},\ \bibinfo
  {pages} {100406} (\bibinfo {year} {2011})}\BibitemShut {NoStop}%
\bibitem [{\citenamefont {Kim}\ \emph {et~al.}(2008)\citenamefont {Kim},
  \citenamefont {Jin}, \citenamefont {Moon}, \citenamefont {Kim}, \citenamefont
  {Park}, \citenamefont {Leem}, \citenamefont {Yu}, \citenamefont {Noh},
  \citenamefont {Kim}, \citenamefont {Oh}, \citenamefont {Park}, \citenamefont
  {Durairaj}, \citenamefont {Cao},\ and\ \citenamefont
  {Rotenberg}}]{IrO_mott_kim_08}%
  \BibitemOpen
  \bibfield  {author} {\bibinfo {author} {\bibfnamefont {B.~J.}\ \bibnamefont
  {Kim}}, \bibinfo {author} {\bibfnamefont {H.}~\bibnamefont {Jin}}, \bibinfo
  {author} {\bibfnamefont {S.~J.}\ \bibnamefont {Moon}}, \bibinfo {author}
  {\bibfnamefont {J.-Y.}\ \bibnamefont {Kim}}, \bibinfo {author} {\bibfnamefont
  {B.-G.}\ \bibnamefont {Park}}, \bibinfo {author} {\bibfnamefont {C.~S.}\
  \bibnamefont {Leem}}, \bibinfo {author} {\bibfnamefont {J.}~\bibnamefont
  {Yu}}, \bibinfo {author} {\bibfnamefont {T.~W.}\ \bibnamefont {Noh}},
  \bibinfo {author} {\bibfnamefont {C.}~\bibnamefont {Kim}}, \bibinfo {author}
  {\bibfnamefont {S.-J.}\ \bibnamefont {Oh}}, \bibinfo {author} {\bibfnamefont
  {J.-H.}\ \bibnamefont {Park}}, \bibinfo {author} {\bibfnamefont
  {V.}~\bibnamefont {Durairaj}}, \bibinfo {author} {\bibfnamefont
  {G.}~\bibnamefont {Cao}}, \ and\ \bibinfo {author} {\bibfnamefont
  {E.}~\bibnamefont {Rotenberg}},\ }\href {\doibase
  10.1103/PhysRevLett.101.076402} {\bibfield  {journal} {\bibinfo  {journal}
  {Phys. Rev. Lett.}\ }\textbf {\bibinfo {volume} {101}},\ \bibinfo {pages}
  {076402} (\bibinfo {year} {2008})}\BibitemShut {NoStop}%
\bibitem [{\citenamefont {Kim}\ \emph {et~al.}(2009)\citenamefont {Kim},
  \citenamefont {Ohsumi}, \citenamefont {Komesu}, \citenamefont {Sakai},
  \citenamefont {Morita}, \citenamefont {Takagi},\ and\ \citenamefont
  {Arima}}]{IrO_mott_arima_09}%
  \BibitemOpen
  \bibfield  {author} {\bibinfo {author} {\bibfnamefont {B.~J.}\ \bibnamefont
  {Kim}}, \bibinfo {author} {\bibfnamefont {H.}~\bibnamefont {Ohsumi}},
  \bibinfo {author} {\bibfnamefont {T.}~\bibnamefont {Komesu}}, \bibinfo
  {author} {\bibfnamefont {S.}~\bibnamefont {Sakai}}, \bibinfo {author}
  {\bibfnamefont {T.}~\bibnamefont {Morita}}, \bibinfo {author} {\bibfnamefont
  {H.}~\bibnamefont {Takagi}}, \ and\ \bibinfo {author} {\bibfnamefont
  {T.}~\bibnamefont {Arima}},\ }\href {\doibase 10.1126/science.1167106}
  {\bibfield  {journal} {\bibinfo  {journal} {Science}\ }\textbf {\bibinfo
  {volume} {323}},\ \bibinfo {pages} {1329} (\bibinfo {year}
  {2009})}\BibitemShut {NoStop}%
\bibitem [{\citenamefont {Kim}\ \emph {et~al.}()\citenamefont {Kim},
  \citenamefont {Casa}, \citenamefont {Upton}, \citenamefont {Gog},
  \citenamefont {Kim}, \citenamefont {Mitchell}, \citenamefont {van
  Veenendaal}, \citenamefont {Daghofer}, \citenamefont {van~den Brink},
  \citenamefont {Khaliullin},\ and\ \citenamefont {Kim}}]{IrO_rixs_kim_11}%
  \BibitemOpen
  \bibfield  {author} {\bibinfo {author} {\bibfnamefont {J.}~\bibnamefont
  {Kim}}, \bibinfo {author} {\bibfnamefont {D.}~\bibnamefont {Casa}}, \bibinfo
  {author} {\bibfnamefont {M.~H.}\ \bibnamefont {Upton}}, \bibinfo {author}
  {\bibfnamefont {T.}~\bibnamefont {Gog}}, \bibinfo {author} {\bibfnamefont
  {Y.-J.}\ \bibnamefont {Kim}}, \bibinfo {author} {\bibfnamefont {J.~F.}\
  \bibnamefont {Mitchell}}, \bibinfo {author} {\bibfnamefont {M.}~\bibnamefont
  {van Veenendaal}}, \bibinfo {author} {\bibfnamefont {M.}~\bibnamefont
  {Daghofer}}, \bibinfo {author} {\bibfnamefont {J.}~\bibnamefont {van~den
  Brink}}, \bibinfo {author} {\bibfnamefont {G.}~\bibnamefont {Khaliullin}}, \
  and\ \bibinfo {author} {\bibfnamefont {B.~J.}\ \bibnamefont {Kim}},\
  }\href@noop {} {\bibinfo  {journal} {arXiv:1110.0759v1 (unpublished)}\
  }\BibitemShut {NoStop}%
\bibitem [{\citenamefont {Wang}\ and\ \citenamefont
  {Senthil}(2011)}]{IrO_HTC_senthil_11}%
  \BibitemOpen
\bibfield  {journal} {  }\bibfield  {author} {\bibinfo {author} {\bibfnamefont
  {F.}~\bibnamefont {Wang}}\ and\ \bibinfo {author} {\bibfnamefont
  {T.}~\bibnamefont {Senthil}},\ }\href {\doibase
  10.1103/PhysRevLett.106.136402} {\bibfield  {journal} {\bibinfo  {journal}
  {Phys. Rev. Lett.}\ }\textbf {\bibinfo {volume} {106}},\ \bibinfo {pages}
  {136402} (\bibinfo {year} {2011})}\BibitemShut {NoStop}%
\bibitem [{\citenamefont {Watanabe}\ \emph {et~al.}(2010)\citenamefont
  {Watanabe}, \citenamefont {Shirakawa},\ and\ \citenamefont
  {Yunoki}}]{IrO_mott_yunoki_10}%
  \BibitemOpen
  \bibfield  {author} {\bibinfo {author} {\bibfnamefont {H.}~\bibnamefont
  {Watanabe}}, \bibinfo {author} {\bibfnamefont {T.}~\bibnamefont {Shirakawa}},
  \ and\ \bibinfo {author} {\bibfnamefont {S.}~\bibnamefont {Yunoki}},\ }\href
  {\doibase 10.1103/PhysRevLett.105.216410} {\bibfield  {journal} {\bibinfo
  {journal} {Phys. Rev. Lett.}\ }\textbf {\bibinfo {volume} {105}},\ \bibinfo
  {pages} {216410} (\bibinfo {year} {2010})}\BibitemShut {NoStop}%
\bibitem [{\citenamefont {{Arita}}\ \emph {et~al.}()\citenamefont {{Arita}},
  \citenamefont {{Kune{\v s}}}, \citenamefont {{Kozhevnikov}}, \citenamefont
  {{Eguiluz}},\ and\ \citenamefont {{Imada}}}]{IrO_mott_arita_11}%
  \BibitemOpen
  \bibfield  {author} {\bibinfo {author} {\bibfnamefont {R.}~\bibnamefont
  {{Arita}}}, \bibinfo {author} {\bibfnamefont {J.}~\bibnamefont {{Kune{\v
  s}}}}, \bibinfo {author} {\bibfnamefont {A.~V.}\ \bibnamefont
  {{Kozhevnikov}}}, \bibinfo {author} {\bibfnamefont {A.~G.}\ \bibnamefont
  {{Eguiluz}}}, \ and\ \bibinfo {author} {\bibfnamefont {M.}~\bibnamefont
  {{Imada}}},\ }\href@noop {} {\bibinfo  {journal} {arXiv:1107.0835
  (unpublished)}\ }\BibitemShut {NoStop}%
\bibitem [{\citenamefont {Jin}\ \emph {et~al.}(2009)\citenamefont {Jin},
  \citenamefont {Jeong}, \citenamefont {Ozaki},\ and\ \citenamefont
  {Yu}}]{IrO_mott_jin_09}%
  \BibitemOpen
\bibfield  {journal} {  }\bibfield  {author} {\bibinfo {author} {\bibfnamefont
  {H.}~\bibnamefont {Jin}}, \bibinfo {author} {\bibfnamefont {H.}~\bibnamefont
  {Jeong}}, \bibinfo {author} {\bibfnamefont {T.}~\bibnamefont {Ozaki}}, \ and\
  \bibinfo {author} {\bibfnamefont {J.}~\bibnamefont {Yu}},\ }\href {\doibase
  10.1103/PhysRevB.80.075112} {\bibfield  {journal} {\bibinfo  {journal} {Phys.
  Rev. B}\ }\textbf {\bibinfo {volume} {80}},\ \bibinfo {pages} {075112}
  (\bibinfo {year} {2009})}\BibitemShut {NoStop}%
\bibitem [{\citenamefont {Mu\~noz}\ \emph {et~al.}(2000)\citenamefont
  {Mu\~noz}, \citenamefont {Illas},\ and\ \citenamefont
  {de~P.~R.~Moreira}}]{CuO_J_Illas_00}%
  \BibitemOpen
  \bibfield  {author} {\bibinfo {author} {\bibfnamefont {D.}~\bibnamefont
  {Mu\~noz}}, \bibinfo {author} {\bibfnamefont {F.}~\bibnamefont {Illas}}, \
  and\ \bibinfo {author} {\bibfnamefont {I.}~\bibnamefont {de~P.~R.~Moreira}},\
  }\href {\doibase 10.1103/PhysRevLett.84.1579} {\bibfield  {journal} {\bibinfo
   {journal} {Phys. Rev. Lett.}\ }\textbf {\bibinfo {volume} {84}},\ \bibinfo
  {pages} {1579} (\bibinfo {year} {2000})}\BibitemShut {NoStop}%
\bibitem [{\citenamefont {Hozoi}\ \emph {et~al.}(2011)\citenamefont {Hozoi},
  \citenamefont {Siurakshina}, \citenamefont {Fulde},\ and\ \citenamefont
  {van~den Brink}}]{CuO_dd_hozoi_11}%
  \BibitemOpen
  \bibfield  {author} {\bibinfo {author} {\bibfnamefont {L.}~\bibnamefont
  {Hozoi}}, \bibinfo {author} {\bibfnamefont {L.}~\bibnamefont {Siurakshina}},
  \bibinfo {author} {\bibfnamefont {P.}~\bibnamefont {Fulde}}, \ and\ \bibinfo
  {author} {\bibfnamefont {J.}~\bibnamefont {van~den Brink}},\ }\href@noop {}
  {\bibfield  {journal} {\bibinfo  {journal} {Nature Scientific Reports}\
  }\textbf {\bibinfo {volume} {1}},\ \bibinfo {pages} {65} (\bibinfo {year}
  {2011})}\BibitemShut {NoStop}%
\bibitem [{\citenamefont {Hozoi}\ and\ \citenamefont
  {Fulde}(2011)}]{reimers_hozoi_fulde}%
  \BibitemOpen
  \bibfield  {author} {\bibinfo {author} {\bibfnamefont {L.}~\bibnamefont
  {Hozoi}}\ and\ \bibinfo {author} {\bibfnamefont {P.}~\bibnamefont {Fulde}},\
  }in\ \href@noop {} {\emph {\bibinfo {booktitle} {Computational Methods for
  Large Systems: Electronic Structre Approaches for Biotechnology and
  Nanotechnology}}},\ \bibinfo {editor} {edited by\ \bibinfo {editor}
  {\bibfnamefont {J.~R.}\ \bibnamefont {Reimers}}}\ (\bibinfo  {publisher}
  {John Wiley \& Sons, Hoboken},\ \bibinfo {year} {2011})\ Chap.~\bibinfo
  {chapter} {6}\BibitemShut {NoStop}%
\bibitem [{\citenamefont {{Crawford}}\ \emph {et~al.}(1994)\citenamefont
  {{Crawford}}, \citenamefont {{Subramanian}}, \citenamefont {{Harlow}},
  \citenamefont {{Fernandez-Baca}}, \citenamefont {{Wang}},\ and\ \citenamefont
  {{Johnston}}}]{IrO_str214_crawford94}%
  \BibitemOpen
  \bibfield  {author} {\bibinfo {author} {\bibfnamefont {M.~K.}\ \bibnamefont
  {{Crawford}}}, \bibinfo {author} {\bibfnamefont {M.~A.}\ \bibnamefont
  {{Subramanian}}}, \bibinfo {author} {\bibfnamefont {R.~L.}\ \bibnamefont
  {{Harlow}}}, \bibinfo {author} {\bibfnamefont {J.~A.}\ \bibnamefont
  {{Fernandez-Baca}}}, \bibinfo {author} {\bibfnamefont {Z.~R.}\ \bibnamefont
  {{Wang}}}, \ and\ \bibinfo {author} {\bibfnamefont {D.~C.}\ \bibnamefont
  {{Johnston}}},\ }\href {\doibase 10.1103/PhysRevB.49.9198} {\bibfield
  {journal} {\bibinfo  {journal} {\prb}\ }\textbf {\bibinfo {volume} {49}},\
  \bibinfo {pages} {9198} (\bibinfo {year} {1994})}\BibitemShut {NoStop}%
\bibitem [{\citenamefont {Okabe}\ \emph {et~al.}(2011)\citenamefont {Okabe},
  \citenamefont {Isobe}, \citenamefont {Takayama-Muromachi}, \citenamefont
  {Koda}, \citenamefont {Takeshita}, \citenamefont {Hiraishi}, \citenamefont
  {Miyazaki}, \citenamefont {Kadono}, \citenamefont {Miyake},\ and\
  \citenamefont {Akimitsu}}]{IrO_str214_okabe11}%
  \BibitemOpen
  \bibfield  {author} {\bibinfo {author} {\bibfnamefont {H.}~\bibnamefont
  {Okabe}}, \bibinfo {author} {\bibfnamefont {M.}~\bibnamefont {Isobe}},
  \bibinfo {author} {\bibfnamefont {E.}~\bibnamefont {Takayama-Muromachi}},
  \bibinfo {author} {\bibfnamefont {A.}~\bibnamefont {Koda}}, \bibinfo {author}
  {\bibfnamefont {S.}~\bibnamefont {Takeshita}}, \bibinfo {author}
  {\bibfnamefont {M.}~\bibnamefont {Hiraishi}}, \bibinfo {author}
  {\bibfnamefont {M.}~\bibnamefont {Miyazaki}}, \bibinfo {author}
  {\bibfnamefont {R.}~\bibnamefont {Kadono}}, \bibinfo {author} {\bibfnamefont
  {Y.}~\bibnamefont {Miyake}}, \ and\ \bibinfo {author} {\bibfnamefont
  {J.}~\bibnamefont {Akimitsu}},\ }\href {\doibase 10.1103/PhysRevB.83.155118}
  {\bibfield  {journal} {\bibinfo  {journal} {Phys. Rev. B}\ }\textbf {\bibinfo
  {volume} {83}},\ \bibinfo {pages} {155118} (\bibinfo {year}
  {2011})}\BibitemShut {NoStop}%
\bibitem [{\citenamefont {Helgaker}\ \emph {et~al.}(2000)\citenamefont
  {Helgaker}, \citenamefont {J{\o}rgensen},\ and\ \citenamefont
  {Olsen}}]{book_QC_00}%
  \BibitemOpen
  \bibfield  {author} {\bibinfo {author} {\bibfnamefont {T.}~\bibnamefont
  {Helgaker}}, \bibinfo {author} {\bibfnamefont {P.}~\bibnamefont
  {J{\o}rgensen}}, \ and\ \bibinfo {author} {\bibfnamefont {J.}~\bibnamefont
  {Olsen}},\ }\href@noop {} {\emph {\bibinfo {title} {Molecular
  Electronic-Structure Theory}}}\ (\bibinfo  {publisher} {Wiley, Chichester},\
  \bibinfo {year} {2000})\BibitemShut {NoStop}%
\bibitem [{\citenamefont {Werner}\ and\ \citenamefont {Knowles}(1988)}]{WK88}%
  \BibitemOpen
  \bibfield  {author} {\bibinfo {author} {\bibfnamefont {H.-J.}\ \bibnamefont
  {Werner}}\ and\ \bibinfo {author} {\bibfnamefont {P.~J.}\ \bibnamefont
  {Knowles}},\ }\href {\doibase 10.1063/1.455556} {\bibfield  {journal}
  {\bibinfo  {journal} {J. Chem. Phys.}\ }\textbf {\bibinfo {volume} {89}},\
  \bibinfo {pages} {5803} (\bibinfo {year} {1988})}\BibitemShut {NoStop}%
\bibitem [{\citenamefont {Berning}\ \emph {et~al.}(2000)\citenamefont
  {Berning}, \citenamefont {Schweizer}, \citenamefont {Werner}, \citenamefont
  {Knowles},\ and\ \citenamefont {Palmieri}}]{SOC_molpro}%
  \BibitemOpen
  \bibfield  {author} {\bibinfo {author} {\bibfnamefont {A.}~\bibnamefont
  {Berning}}, \bibinfo {author} {\bibfnamefont {M.}~\bibnamefont {Schweizer}},
  \bibinfo {author} {\bibfnamefont {H.-J.}\ \bibnamefont {Werner}}, \bibinfo
  {author} {\bibfnamefont {P.~J.}\ \bibnamefont {Knowles}}, \ and\ \bibinfo
  {author} {\bibfnamefont {P.}~\bibnamefont {Palmieri}},\ }\href {\doibase
  10.1080/00268970009483386} {\bibfield  {journal} {\bibinfo  {journal} {Mol.
  Phys.}\ }\textbf {\bibinfo {volume} {98}},\ \bibinfo {pages} {1823} (\bibinfo
  {year} {2000})}\BibitemShut {NoStop}%
\bibitem [{\citenamefont {Werner}\ \emph {et~al.}()\citenamefont {Werner},
  \citenamefont {Knowles}, \citenamefont {Knizia}, \citenamefont {Manby},
  \citenamefont {{Sch\"{u}tz}} \emph {et~al.}}]{molpro_brief}%
  \BibitemOpen
  \bibfield  {author} {\bibinfo {author} {\bibfnamefont {H.-J.}\ \bibnamefont
  {Werner}}, \bibinfo {author} {\bibfnamefont {P.~J.}\ \bibnamefont {Knowles}},
  \bibinfo {author} {\bibfnamefont {G.}~\bibnamefont {Knizia}}, \bibinfo
  {author} {\bibfnamefont {F.~R.}\ \bibnamefont {Manby}}, \bibinfo {author}
  {\bibfnamefont {M.}~\bibnamefont {{Sch\"{u}tz}}}  \emph {et~al.},\
  }\href@noop {} {}\bibinfo {note} {{\sc molpro} 2010, see
  http://www.molpro.net}\BibitemShut {NoStop}%
\bibitem [{\citenamefont {Figgen}\ \emph {et~al.}(2009)\citenamefont {Figgen},
  \citenamefont {Peterson}, \citenamefont {Dolg},\ and\ \citenamefont
  {Stoll}}]{ECP_Stoll_2}%
  \BibitemOpen
  \bibfield  {author} {\bibinfo {author} {\bibfnamefont {D.}~\bibnamefont
  {Figgen}}, \bibinfo {author} {\bibfnamefont {K.~A.}\ \bibnamefont
  {Peterson}}, \bibinfo {author} {\bibfnamefont {M.}~\bibnamefont {Dolg}}, \
  and\ \bibinfo {author} {\bibfnamefont {H.}~\bibnamefont {Stoll}},\ }\href
  {\doibase 10.1063/1.3119665} {\bibfield  {journal} {\bibinfo  {journal} {J.
  Chem. Phys.}\ }\textbf {\bibinfo {volume} {130}},\ \bibinfo {eid} {164108}
  (\bibinfo {year} {2009})}\BibitemShut {NoStop}%
\bibitem [{\citenamefont {Fuentealba}\ \emph {et~al.}(1985)\citenamefont
  {Fuentealba}, \citenamefont {von Szentpaly}, \citenamefont {Preuss},\ and\
  \citenamefont {Stoll}}]{ECP_Stoll_1}%
  \BibitemOpen
  \bibfield  {author} {\bibinfo {author} {\bibfnamefont {P.}~\bibnamefont
  {Fuentealba}}, \bibinfo {author} {\bibfnamefont {L.}~\bibnamefont {von
  Szentpaly}}, \bibinfo {author} {\bibfnamefont {H.}~\bibnamefont {Preuss}}, \
  and\ \bibinfo {author} {\bibfnamefont {H.}~\bibnamefont {Stoll}},\ }\href
  {http://stacks.iop.org/0022-3700/18/i=7/a=010} {\bibfield  {journal}
  {\bibinfo  {journal} {J. Phys. B}\ }\textbf {\bibinfo {volume} {18}},\
  \bibinfo {pages} {1287} (\bibinfo {year} {1985})}\BibitemShut {NoStop}%
\bibitem [{\citenamefont {Ament}\ \emph {et~al.}(2011)\citenamefont {Ament},
  \citenamefont {Khaliullin},\ and\ \citenamefont {van~den
  Brink}}]{IrO_rixs_ament_11}%
  \BibitemOpen
  \bibfield  {author} {\bibinfo {author} {\bibfnamefont {L.~J.~P.}\
  \bibnamefont {Ament}}, \bibinfo {author} {\bibfnamefont {G.}~\bibnamefont
  {Khaliullin}}, \ and\ \bibinfo {author} {\bibfnamefont {J.}~\bibnamefont
  {van~den Brink}},\ }\href {\doibase 10.1103/PhysRevB.84.020403} {\bibfield
  {journal} {\bibinfo  {journal} {Phys. Rev. B}\ }\textbf {\bibinfo {volume}
  {84}},\ \bibinfo {pages} {020403} (\bibinfo {year} {2011})}\BibitemShut
  {NoStop}%
\bibitem [{\citenamefont {Kimchi}\ and\ \citenamefont
  {You}(2011)}]{IrO_kitaev_heisenberg_kimchi_11}%
  \BibitemOpen
  \bibfield  {author} {\bibinfo {author} {\bibfnamefont {I.}~\bibnamefont
  {Kimchi}}\ and\ \bibinfo {author} {\bibfnamefont {Y.-Z.}\ \bibnamefont
  {You}},\ }\href {\doibase 10.1103/PhysRevB.84.180407} {\bibfield  {journal}
  {\bibinfo  {journal} {Phys. Rev. B}\ }\textbf {\bibinfo {volume} {84}},\
  \bibinfo {pages} {180407} (\bibinfo {year} {2011})}\BibitemShut {NoStop}%
\bibitem [{\citenamefont {Bhattacharjee}\ \emph {et~al.}()\citenamefont
  {Bhattacharjee}, \citenamefont {Lee},\ and\ \citenamefont
  {Kim}}]{NaIrO_mgn_kim_11}%
  \BibitemOpen
  \bibfield  {author} {\bibinfo {author} {\bibfnamefont {S.}~\bibnamefont
  {Bhattacharjee}}, \bibinfo {author} {\bibfnamefont {S.-S.}\ \bibnamefont
  {Lee}}, \ and\ \bibinfo {author} {\bibfnamefont {Y.~B.}\ \bibnamefont
  {Kim}},\ }\href@noop {} {\bibinfo  {journal} {arXiv:1108.1806v2
  (unpublished)}\ }\BibitemShut {NoStop}%
\bibitem [{\citenamefont {Jin}\ \emph {et~al.}()\citenamefont {Jin},
  \citenamefont {Kim}, \citenamefont {Jeong}, \citenamefont {Kim},\ and\
  \citenamefont {Yu}}]{NaIrO_mgn_yu_11}%
  \BibitemOpen
\bibfield  {journal} {  }\bibfield  {author} {\bibinfo {author} {\bibfnamefont
  {H.}~\bibnamefont {Jin}}, \bibinfo {author} {\bibfnamefont {H.}~\bibnamefont
  {Kim}}, \bibinfo {author} {\bibfnamefont {H.}~\bibnamefont {Jeong}}, \bibinfo
  {author} {\bibfnamefont {C.~H.}\ \bibnamefont {Kim}}, \ and\ \bibinfo
  {author} {\bibfnamefont {J.}~\bibnamefont {Yu}},\ }\href@noop {} {\bibinfo
  {journal} {arXiv:0907.0743v1 (unpublished)}\ }\BibitemShut {NoStop}%
\bibitem [{foo()}]{footnote_SO_96}%
  \BibitemOpen
\bibfield  {journal} {  }\bibinfo {note} {There are another 31 spin-orbit
  states in an interval of $\approx80$ meV above the GS, related to the
  different possibilities of coupling the effective $j\!=\!1/2$ moments of
  those five Ir sites.}\BibitemShut {Stop}%
\bibitem [{\citenamefont {Andlauer}\ \emph {et~al.}(1976)\citenamefont
  {Andlauer}, \citenamefont {Schneider},\ and\ \citenamefont
  {Tolksdorf}}]{Andlauer76}%
  \BibitemOpen
  \bibfield  {author} {\bibinfo {author} {\bibfnamefont {B.}~\bibnamefont
  {Andlauer}}, \bibinfo {author} {\bibfnamefont {J.}~\bibnamefont {Schneider}},
  \ and\ \bibinfo {author} {\bibfnamefont {W.}~\bibnamefont {Tolksdorf}},\
  }\href@noop {} {\bibfield  {journal} {\bibinfo  {journal} {Phys. Stat. Sol.
  B}\ }\textbf {\bibinfo {volume} {73}},\ \bibinfo {pages} {533} (\bibinfo
  {year} {1976})}\BibitemShut {NoStop}%
\bibitem [{\citenamefont {Su}\ \emph {et~al.}(2009)\citenamefont {Su},
  \citenamefont {He},\ and\ \citenamefont {Zheng}}]{Su09}%
  \BibitemOpen
  \bibfield  {author} {\bibinfo {author} {\bibfnamefont {P.}~\bibnamefont
  {Su}}, \bibinfo {author} {\bibfnamefont {L.}~\bibnamefont {He}}, \ and\
  \bibinfo {author} {\bibfnamefont {W.-C.}\ \bibnamefont {Zheng}},\ }\href@noop
  {} {\bibfield  {journal} {\bibinfo  {journal} {J. Alloys Comp.}\ }\textbf
  {\bibinfo {volume} {474}},\ \bibinfo {pages} {31} (\bibinfo {year}
  {2009})}\BibitemShut {NoStop}%
\bibitem [{\citenamefont {Ishii}\ \emph {et~al.}(2011)\citenamefont {Ishii},
  \citenamefont {Jarrige}, \citenamefont {Yoshida}, \citenamefont {Ikeuchi},
  \citenamefont {Mizuki}, \citenamefont {Ohashi}, \citenamefont {Takayama},
  \citenamefont {Matsuno},\ and\ \citenamefont {Takagi}}]{IrO_rixs_ishii_11}%
  \BibitemOpen
  \bibfield  {author} {\bibinfo {author} {\bibfnamefont {K.}~\bibnamefont
  {Ishii}}, \bibinfo {author} {\bibfnamefont {I.}~\bibnamefont {Jarrige}},
  \bibinfo {author} {\bibfnamefont {M.}~\bibnamefont {Yoshida}}, \bibinfo
  {author} {\bibfnamefont {K.}~\bibnamefont {Ikeuchi}}, \bibinfo {author}
  {\bibfnamefont {J.}~\bibnamefont {Mizuki}}, \bibinfo {author} {\bibfnamefont
  {K.}~\bibnamefont {Ohashi}}, \bibinfo {author} {\bibfnamefont
  {T.}~\bibnamefont {Takayama}}, \bibinfo {author} {\bibfnamefont
  {J.}~\bibnamefont {Matsuno}}, \ and\ \bibinfo {author} {\bibfnamefont
  {H.}~\bibnamefont {Takagi}},\ }\href {\doibase 10.1103/PhysRevB.83.115121}
  {\bibfield  {journal} {\bibinfo  {journal} {Phys. Rev. B}\ }\textbf {\bibinfo
  {volume} {83}},\ \bibinfo {pages} {115121} (\bibinfo {year}
  {2011})}\BibitemShut {NoStop}%
\bibitem [{\citenamefont {Moon}\ \emph {et~al.}(2006)\citenamefont {Moon},
  \citenamefont {Kim}, \citenamefont {Kim}, \citenamefont {Lee}, \citenamefont
  {Kim}, \citenamefont {Park}, \citenamefont {Kim}, \citenamefont {Oh},
  \citenamefont {Nakatsuji}, \citenamefont {Maeno}, \citenamefont {Nagai},
  \citenamefont {Ikeda}, \citenamefont {Cao},\ and\ \citenamefont
  {Noh}}]{IrO_optics_moon_06}%
  \BibitemOpen
  \bibfield  {author} {\bibinfo {author} {\bibfnamefont {S.~J.}\ \bibnamefont
  {Moon}}, \bibinfo {author} {\bibfnamefont {M.~W.}\ \bibnamefont {Kim}},
  \bibinfo {author} {\bibfnamefont {K.~W.}\ \bibnamefont {Kim}}, \bibinfo
  {author} {\bibfnamefont {Y.~S.}\ \bibnamefont {Lee}}, \bibinfo {author}
  {\bibfnamefont {J.-Y.}\ \bibnamefont {Kim}}, \bibinfo {author} {\bibfnamefont
  {J.-H.}\ \bibnamefont {Park}}, \bibinfo {author} {\bibfnamefont {B.~J.}\
  \bibnamefont {Kim}}, \bibinfo {author} {\bibfnamefont {S.-J.}\ \bibnamefont
  {Oh}}, \bibinfo {author} {\bibfnamefont {S.}~\bibnamefont {Nakatsuji}},
  \bibinfo {author} {\bibfnamefont {Y.}~\bibnamefont {Maeno}}, \bibinfo
  {author} {\bibfnamefont {I.}~\bibnamefont {Nagai}}, \bibinfo {author}
  {\bibfnamefont {S.~I.}\ \bibnamefont {Ikeda}}, \bibinfo {author}
  {\bibfnamefont {G.}~\bibnamefont {Cao}}, \ and\ \bibinfo {author}
  {\bibfnamefont {T.~W.}\ \bibnamefont {Noh}},\ }\href {\doibase
  10.1103/PhysRevB.74.113104} {\bibfield  {journal} {\bibinfo  {journal} {Phys.
  Rev. B}\ }\textbf {\bibinfo {volume} {74}},\ \bibinfo {pages} {113104}
  (\bibinfo {year} {2006})}\BibitemShut {NoStop}%
\bibitem [{\citenamefont {Broer}\ \emph {et~al.}(2003)\citenamefont {Broer},
  \citenamefont {Hozoi},\ and\ \citenamefont {Nieuwpoort}}]{CuO_J_NOCI_03}%
  \BibitemOpen
  \bibfield  {author} {\bibinfo {author} {\bibfnamefont {R.}~\bibnamefont
  {Broer}}, \bibinfo {author} {\bibfnamefont {L.}~\bibnamefont {Hozoi}}, \ and\
  \bibinfo {author} {\bibfnamefont {W.~C.}\ \bibnamefont {Nieuwpoort}},\ }\href
  {\doibase 10.1080/0026897021000035205} {\bibfield  {journal} {\bibinfo
  {journal} {Mol. Phys.}\ }\textbf {\bibinfo {volume} {101}},\ \bibinfo {pages}
  {233} (\bibinfo {year} {2003})}\BibitemShut {NoStop}%
\end{thebibliography}
%

\end{document}